\documentclass[runningheads]{llncs}
\usepackage{graphicx}
\usepackage{threeparttable}
\begin{document}
\title{An Analysis of Classification Approaches for Hit Song Prediction using Engineered Metadata Features with Lyrics and Audio Features}
\titlerunning{An Analysis of Classification Approaches for Hit Song Prediction}
%
\author{Mengyisong Zhao\inst{1}\and
Morgan Harvey\inst{1}\and
David Cameron\inst{1}\and 
Frank Hopfgartner\inst{2}\and
Valerie J. Gillet\inst{1}}
\authorrunning{M. Zhao et al.}
%
\institute{The University of Sheffield, Sheffield S10 2TN, UK
\email{mzhao18@sheffield.ac.uk, m.harvey@sheffield.ac.uk, d.s.cameron@sheffield.ac.uk, v.gillet@sheffield.ac.uk}\\
\and
Universit{\"a}t Koblenz, 56072 Koblenz, Germany
\email{hopfgartner@uni-koblenz.de}}
\maketitle              
\begin{abstract}
Hit song prediction, one of the emerging fields in music information retrieval (MIR), remains a considerable challenge. Being able to understand what makes a given song a hit is clearly beneficial to the whole music industry. Previous approaches to hit song prediction have focused on using audio features of a record. This study aims to improve the prediction result of the top 10 hits among Billboard Hot 100 songs using more alternative metadata, including song audio features provided by Spotify, song lyrics, and novel metadata-based features (title topic, popularity continuity and genre class). Five machine learning approaches are applied, including: k-nearest neighbours, Naïve Bayes, Random Forest, Logistic Regression and Multilayer Perceptron. Our results show that Random Forest (RF) and Logistic Regression (LR) with all features (including novel features, song audio features and lyrics features) outperforms other models, achieving 89.1\% and 87.2\% accuracy, and 0.91 and 0.93 AUC, respectively. Our findings also demonstrate the utility of our novel music metadata features, which contributed most to the models’ discriminative performance.

\end{abstract}
\section{Introduction}
Music labels spend more than \$4.5 billion every year discovering new talented artists and producing popular songs \cite{1 IFPI Global Music Report}. Precipitated by the growing importance of online digital music platforms and recent advancements in machine learning and big data technologies, a new research area called hit song science has attracted increasing attention \cite{2 Greenberg and Rentfrow}. A successful hit song prediction approach could bring considerable benefits to many music lifecycle stakeholders. Early hit song prediction studies illustrate the complexity of this problem, delivering only weak classification results \cite{3 Dhanaraj and Logan,4 Salganik et al.,5 Pachet et al.,6 Herremans et al.}. In recent years, more advanced approaches have been able to accurately predict hits and non-hits using audio features \cite{7 Ni et al.,8 Georgieva et al.,9 Middlebrook et al.,10 Kim and Oh,11 Fan and Casey,12 Song Popularity Predictor Homepage,13 Kawawa-Beaudan and Garza}; however, many other potentially useful sources of information about the songs are also available. In this study, we employ 12 Spotify audio features (energy, liveness, tempo, speechiness, acousticness, time\_signature, key, duration\_ms, loudness, valence, mode and danceability), these features are drawn directly from Spotify, together with \textit{novel features} based on Billboard music metadata (popularity continuity, genre class and title topic), as well as the topics extracted from the songs’ lyrics to identify Top 10 hits among Top 100 hits. To our knowledge, this work is the first attempt to improve hit song prediction by extracting features from the topic of song titles and by using a song’s prior popularity information. We examine the effectiveness of these novel features together with song audio and lyrics features for hit song prediction using a variety of machine learning approaches, including k-nearest neighbours (kNN), Naïve Bayes (NB), Random Forest (RF), Logistic Regression (LR) and Multilayer Perceptron (MLP). Our findings demonstrate the utility of the new features and provide state-of-the-art prediction performance, as well as providing promising avenues for future work in this area.

\section{Related Work}
Hit song prediction (HSP) has been investigated frequently in recent decades. Much seminal work failed to accurately predict hit songs, with some work even suggesting that popularity was not predictable \cite{4 Salganik et al.,5 Pachet et al.}. An early approach by Dhanaraj and Logan \cite{3 Dhanaraj and Logan} achieved promising results by using a SVM model to classify top 1 songs through acoustic and lyrics data. However, they provided only scant details about their data gathering, feature engineering, model training and parameter optimization procedure and found textual features to be more predictive than audio analysis. Salganik et al.\cite{4 Salganik et al.}, and Pachet and Roy \cite{5 Pachet et al.} attempted to reproduce Dhanaraj and Logan's work but failed to achieve a similar level of accuracy.

Various algorithms have been applied to tackle this task, among them: Logistic Regression (LR), Support Vector Machine (SVM) and Neural Networks (NN) are commonly used \cite{3 Dhanaraj and Logan,6 Herremans et al.,8 Georgieva et al.,11 Fan and Casey}.  Ni et al. \cite{7 Ni et al.} gained promising results in predicting UK Top 5 hits on the Top 40 single song charts, but again little implementation detail was provided. Fan and Casey \cite{11 Fan and Casey} used LR and SVM models to predict British and Chinese hit songs but found that audio features worked better for predicting Chinese hits than British ones, and that textual features worked best overall. 

Herremans et al. \cite{6 Herremans et al.} focussed particularly on dance songs and classified hits using five machine learning models. Their research affirmed the importance of audio features; however, they achieved relatively poor accuracy results, perhaps due to their use of a large number of features without performing any feature selection. Georgieva et al. \cite{8 Georgieva et al.} compared six machine-learning algorithms when conducting Billboard hit song prediction; the most successful algorithms were LR and a NN with a single hidden layer. Their work also demonstrated the utility of Spotify’s audio features for this task. Nasreldin \cite{12 Song Popularity Predictor Homepage} did similar research but identified XGBoost as the top performing classifier; in their study the SVM model performed the worst. As they only use the raw data without any feature selection, they only achieved accuracy results similar to those of Herremans et al. \cite{6 Herremans et al.}

Recently, Zangerle et al. \cite{15 Zangerle et al.} adopted deep neural networks and treat HSP as a regression task, and their experimental results show that the wide and deep neural network-based approach performed best, achieving 72.04\% accuracy. However, the common problem with deep neural networks is that their results are hard to interpret. Essa et al. \cite{16 Essa et al.} tried to solve the HIS task by using both classification and regression models. They considered audio features alone and, through adopting seven machine learning models, they achieved results suggesting that both machine learning approaches (classification and regression) can be used for HSP.  

Although previous studies have made a large contribution to this topic, it is still unclear which features can be used to successfully classify hit songs when including audio features, music metadata and song lyrics, and in what combination. Audio features have shown promise, but only raw terms have been used to construct features to date \cite{5 Pachet et al.,6 Herremans et al.,7 Ni et al.}. Textual features have rarely been adopted in hit song prediction tasks and, although Singhi and Brown \cite{17 Singhi and Brown} did attempt to extract 31 song lyrics features and build SVM model to predict hit songs, the performance achieved was not inspiring. 

\section{Data and Methodology}
\subsection{Data collection and preprocessing}
To investigate hit song prediction, we obtained Billboard hot 100 songs data from the open-source platform \textit{data.world} named ``\textit{Billboard Hot-100 Songs 2000-2018 w/Spotify Data+Lyric}''\footnote{data.world/typhon/billboard-hot-100-songs-2000-2018-w-spotify-data-lyrics} . The dataset includes all songs in the Billboard hot 100 weekly charts from 2007 to 2017, as well as audio features, metadata and lyrics of each song provided by Spotify. The raw dataset includes 33 attributes in total. We firstly remove the irrelevant features (e.g., spotify\_link, video\_link, analysis\_url). Then, we define “hits'' in this context to be songs whose highest position in the Billboard Hot 100 list was at rank 10 or above to produce a binary classification of “hit” (1) that at some point reached the Top 10 or “not-hit” (0) that never reached the Top 10. The features used in this study include those engineered based on metadata (e.g., weeks, song title, music genre), 12 Spotify audio features, as well as lyrics of each song. 273 songs had missing audio features data and/or lyrics, and were subsequently removed as it would not be possible to extrapolate or estimate such features. This left 3581 unique songs in the final data set: 507 hits and 3074 non-hits.

\subsection{Feature Engineering} 
We engineered several additional features to augment the existing metadata features from the original Billboard data and the Spotify audio features. \textit{Popularity continuity} was created to represent the sum of each song’s popular duration (i.e., how many weeks it had already been listed in the hot 100 chart prior to the week of interest). Songs already present in the chart for more than 50 weeks were assigned a 3; those present for between 20 and 50 were assigned to 2; those between 10 and 20 were assigned 1; otherwise, a song was assigned 0. Unlike classical music, popular music has relatively rapid iterations \cite{19 Kinga}. The majority of songs only remain in the chart for a short period, typically less than 20 weeks. Therefore, we assign a number based on 3 duration splits where the assigned numbers are only based on weekly duration data. The \textit{song title topic} feature was created based on the song title. We removed symbols, punctuation, short terms (i.e., fewer than 4 chars) and stopwords from the data, then, inspired by \cite{3 Dhanaraj and Logan}, used a bag\-of\-words representation with Latent Dirichlet Allocation (LDA) to extract topics from the song titles. In total, ten topics were extracted, and each song was assigned to the topic number with the highest probability for that song in $\theta$. The numerical variable named \textit{genre class} was created to replace the existing string variable \textit{broad\_genre} in which each genre was assigned a numerical value: 1 to 6 representing country, electronic dance music (edm), pop, r\&b, rock and rap music respectively. We treat song lyrics similar to how we treat song titles, the only difference being the number of topics: 20 topics were extracted from the lyrics. This is because lyrics are far longer than titles, thus providing sufficient data to extract a larger number of more meaningful topics. Each song was assigned to the topic number with the highest probability for that song in $\theta$.

\subsection{Training Environment} 
Min-max normalization method was applied to accelerate the algorithm convergence speed \cite{16 Essa et al.}. After preprocessing and feature engineering, a total of 16 features were used for model building. We treat each song as an individual, temporal factors were not considered in our experiment, the data were split into training and testing sets using a ratio of 80:20 and, due to the relatively small size of the overall data set, 5-fold cross validation was applied instead of an individual validation set. Due to the highly imbalanced classes (i.e., most songs are not top-10 hits), Synthetic Minority Over-Sampling (SMOTE) was adopted inspired by Chawla et al. \cite{18 Chawla et al.}, which could effectively increase the accuracy of minority class (hit song) prediction. In this paper, 5 nearest neighbours have currently used to over-sampling the minority class (hit songs). resulting in a final training set of 4918 songs (including hits 2459 and 2459 non-hits). Forward feature selection was carried out. All models were trained and tested using KNIME 4.4.0\footnote{https://www.knime.com/}.

\subsection{Model Setup and Optimisation}
This study examined five commonly-used machine learning approaches from the prior literature, and all the model parameter tuning has been using 5-fold cross validation. The model hyperparameter and their optimum values are shown in Table~\ref{tab1}.

 \textit{k-Nearest Neighbour (kNN)}. We tested values of k between 1 to 20 to seek for more appropriate neighborhood distance when predicting hit songs, and when we tuned the hyperparameter to k = 1 has achieved most effective accuracy. 
 
 \textit{Naïve Bayes (NB)}. We tested the default probability from 0.001 to 1 every 0.01, the best setting was default probability = 0.031. 
 
 \textit{Random Forest (RF)}. When using RF to train our model, the different split criterion algorithm provides varied performance, which includes information gain, information gain ratio, and Gini index. We tested the number of models of all algorithms from 50 to 1000 every 50, and the Gini index achieved best performance at 600 numbers of models. 
 
 \textit{Logistic Regression (LR)}. We tested four ways to solve the equation, iteratively reweighted least squares with Gauss, iteratively reweighted least squares with Laplace, stochastic average gradient with Gauss, stochastic average gradient with Laplace. We find out using iteratively reweighted least squares with Laplace regularization to solve the equation is more effective. The Laplace equals to 3 has been accepted as best performance.
 
 \textit{Neural Network (NN)}. Multilayer perceptron (MLP) model consisting of an input layer, a hidden layer, and an output layer has been conducted in this study. We tested the Maximum number of iterations from 500 to 5000 with 500 stop sizes, Number of hidden layers and Number of hidden neurons per layer from 1 to 25 was measured every 3. The best parameter tuning result is 4500, 4, 22, respectively. 

\begin{table}
\vspace{-1.5em}
\centering
\caption{All model hyperparameter tuning optimization value.}\label{tab1}
\begin{tabular}{|l|l|l|}
\hline
Classifier & Hyperparameter & Value\\
\hline
kNN &  {K value} & 1\\
NB &  {Default probability} & 0.031\\
RF & {Gini index: number of models} & 600\\
LR & {Laplace} & 3\\
NN & {Maximum number of iterations} & 4500\\
~ & {Number of hidden layers} & 4\\
~ & {Number of hidden neurons per layer} & 22\\
\hline
\end{tabular}
\end{table}
\vspace{-1.5em}

\section{Findings, Results and Limitations}
Our results include an analysis of accuracy, as well as AUC and the number of features used as a measure of parsimony (see Table~\ref{tab2} and Table~\ref{tab3}). We compare models trained using all features, including our novel engineered ones, audio features and lyrics features together, against three “baseline” models, audio features alone, audio features and original metadata features, as well as novel features and audio features model. It is notable that Random Forest (Accuracy=89.1\%, AUC=0.91) and Logistic Regression (Accuracy=87.2\%, AUC=0.93) with all features performed best according to both metrics. Logistic Regression with Laplace regularisation achieves the best AUC score while only using 4 features. According to Han et al. \cite{20 Han et al.}, the reason L1 regularisation is more appropriate to this task could be it capable of reduce the coefficients of some features to zero and generate a spare solution. Random Forest achieved the best accuracy result, but required seven features to train the model, which leads to longer training times, and poorer explainability. MLP shows average performance in this task; this model requires a maximum number of features according to Table~\ref{tab2}, and longest training time to achieve the best result, perhaps because the volume of the data available is insufficient to train the network well. Naïve Bayes performs worst on accuracy, but better on AUC score, which means this model has great ability on identifying hits but weak on identifying non-hits. 

\begin{table}[htbp]
\vspace{-1.5em}
\centering
\caption{Features selected for each model.}
\begin{threeparttable}
\label{tab2}
\resizebox{\linewidth}{!}{ 
\begin{tabular}{|l|l|}
\hline
Classifier & Accepted Feature Combination\\
\hline
kNN &  {\textit{popularity continuity\tnote{1}, song title topic, genre class}, energy, liveness, key, \textbf{lyrics topic}\tnote{2}}\\
NB &  {\textit{popularity continuity, genre class}, key, loudness}\\
RF & {\textit{popularity continuity, genre class, song title topic}, key, valence, energy, \textbf{lyrics topic}}\\
LR & {\textit{popularity continuity, genre class}, \textbf{lyrics topic}, danceability}\\
NN & {\textit{popularity continuity, genre class}, key, \textit{song title topic}, \textbf{lyrics topic}, acousticness, liveness, tempo, danceability}\\
\hline
\end{tabular}
}
 \begin{tablenotes}
  \footnotesize
  \item[1] Novel features are marked in italics.
  \item[2] Lyrics feature is marked in bold.
 \end{tablenotes}
\end{threeparttable}    
\end{table}
\vspace{-1.5em}

\begin{table}[htbp]
\vspace{-1.5em}
\centering
\caption{All model training and test results summarisation and comparison.}\label{tab3}
\begin{threeparttable}
\resizebox{\linewidth}{!}{ 
\begin{tabular}{|l|l|l|l|l|}
\hline
~ & 5-fold CV Accuracy & 5-fold CV AUC & Model Test Accuracy & Model Test AUC\\
\hline
KNN (Audio) & 83.98\% & 0.847 &	79.92\%	& 0.530\\
KNN (Metadata+audio) & 90.85\% & 0.917 & 82.08\% & 0.748\\
KNN (NFE\tnote{1} +audio) & 93.79\% & 0.930 & 86.05\% & \textbf{0.775}\tnote{2}\\
KNN (NFE+audio+lyrics) & \textbf{94.31\%} & \textbf{0.937} & \textbf{86.38\%} & 0.745\\
NB (Audio) & 62.71\% & 0.697  & 42.82\% & 0.609\\
NB (Metadata+audio) & 82.78\% & 0.915 & 71.13\% & 0.899\\
NB (NFE+audio) & \textbf{86.26\%} & 0.924 & 74.76\% & \textbf{0.922}\\
NB (NFE+audio+lyrics) & 86.23\% & \textbf{0.931} & \textbf{78.52\%} & 0.900\\
RF (Audio) & 79.22\% & 0.876 & 71.13\% & 0.629\\
RF (Metadata+audio) & 91.62\% & 0.977 & 74.76\% & 0.869\\
RF (NFE+audio) & 93.84\% & 0.980 & 87.59\% & 0.908\\
RF (NFE+audio+lyrics) & \textbf{95.1\%} & \textbf{0.989} & \textbf{89.12\%} & \textbf{0.912}\\
LR (Audio) & 61.26\% & 0.649 & 57.88\% & 0.603\\
LR (Metadata+audio) & 84.83\% & 0.917 & 83.54\% & 0.927\\
LR (NFE+audio) & 86.15\% & 0.928 & 86.47\% & 0.923\\
        LR (NFE+audio+lyrics) & \textbf{87.07\%} & \textbf{0.933} & \textbf{87.17\%} & \textbf{0.927}\\
MLP (Audio) & 68.20\% & 0.756 & 63.60\% & 0.563\\
MLP (Metadata+audio) & 87.48\% & 0.923 & 76.85\% & \textbf{0.847}\\
MLP (NFE+audio) & 88.0\% & 0.929 & 79.36\% & 0.734\\
MLP (NFE+audio+lyrics) & \textbf{90.04\%} & \textbf{0.931} & \textbf{84.66\%} & 0.808\\
\hline
\end{tabular}}
    \begin{tablenotes}
  \footnotesize
  \item[1] NFE stands for abbreviation of novel feature engineering.
  \item[2] The best performance has been marked in \textbf{bold}.
 \end{tablenotes}
 \end{threeparttable}
\end{table}
\vspace{-1.5em}

Compared to the baseline method, all the model test accuracy results with our novel metadata features provided significant performance improvement seen in Table~\ref{tab3}, which proved our novel metadata features have contributed impact to HSP task. When adding \textit{song lyrics topic} features, the accuracy score of all models are slightly increased, the AUC score of kNN and NB are decreased for .030 and .022 respectively, probably because the lyrics topic increase the complexity of features, which might be hard for both algorithm to classify the patterns of hits and non-hits. The novel variables almost frequently in the list of automatically selected features as shown in Table~\ref{tab2}, demonstrating their discriminative power. The utility of \textit{popularity continuity} indicates that the longer a song in a particular genre can maintain a position in the charts, the more likely it is to become a hit song. Certain topically-coherent sets of terms, such as \textit{love, girls, life}, and  \textit{hearts} are more likely to appear in the hits than non-hits, as captured in the \textit{song title topic} and  \textit{lyrics topic} feature. Based on the ablation studies, some of the Spotify audio features such as \textit{key, liveness, energy}, and  \textit{danceability} are also important when classifying hit songs but less consistently so than our \textit{novel features} and \textit{song lyrics feature}.  The contributed features are varied between each model. Compared to \textit{song title topic}, \textit{song lyrics feature} shows more contribution when using these two features together to identify hit songs.

The result of this study supports the findings of \cite{6 Herremans et al.,8 Georgieva et al.,13 Kawawa-Beaudan and Garza,14 Borg et al.} that music metadata, audio features and lyrics can be used to classify hit songs through machine learning approaches. Adding all features together has achieved the best performance of all models. Moreover, we have been able to outperform the baseline results of \cite{8 Georgieva et al.,9 Middlebrook et al.,10 Kim and Oh}, as their work achieved an accuracy score around 60\% to 87\% compared to our work, which gave accuracy scores around 79\% to 89\%. 

As future work, we intend to further enrich our models by developing more features based on, for example, music reviews and social tags. More complex and granular genre classifications such as different types of music from various cultures, like Latin music or dance songs from India could be used to extend our model. Furthermore, a larger dataset covering a longer period will be examined. As the hit songs identified in our study can be defined as extremely popular songs (top 10 among100), the model generalisation ability may need more tests, particularly adding songs never achieved in billboard top 100. The substance of a hit song may change over time, and we will consider more complex models that include temporal aspects to model changes in genres and topical popularity over time. Other audio-based features could also be considered, such as Mel-frequency Cepstral Coefficients (MFCC) and compared with the Spotify audio features.

%
%
%

\begin{thebibliography}{8}
\bibitem{1 IFPI Global Music Report}
IFPI Global Music Report Homepage, \url{http://www.ifpicr.cz/ifpi-global-music-report-2016/}. last accessed 21 Nov 2022

\bibitem{2 Greenberg and Rentfrow}
Greenberg, D.M., and Rentfrow, P.J.: Music and big data: a new frontier. Current opinion in behavioral sciences \textbf{18}, 50-56 (2017)

\bibitem{3 Dhanaraj and Logan}
Dhanaraj, R, and Logan, B.: Automatic Prediction of Hit Songs. In ISMIR, 488-491 (2005)

\bibitem{4 Salganik et al.}
Salganik, M.J., Dodds, P.S. and Watts, J.D.: Experimental study of inequality and unpredictability in an artificial cultural market. science 311, 5762, 854-856 (2006)

\bibitem{5 Pachet et al.}
Pachet, F., and Roy, P.: Hit Song Science Is Not Yet a Science. In ISMIR, 355-360 (2008)

\bibitem{6 Herremans et al.}
Herremans, D., Martens, D. and Sörensen, K.: Dance hit song prediction. Journal of New Music Research \textbf{43}(3), 291-302 (2014)

\bibitem{7 Ni et al.}
Ni, Y., Santos-Rodriguez, R., Mcvicar, R. and De Bie, T.: Hit song science once again a science. In 4th International Workshop on Machine Learning and Music (2011)

\bibitem{8 Georgieva et al.}
Georgieva, E., Marcella S. and Nicholas B. Hitpredict.: Predicting hit songs using Spotify data. STANFORD COMPUTER SCIENCE 229: MACHINE LEARNING (2018)

\bibitem{9 Middlebrook et al.}
Middlebrook, K. and Sheik, K.: Song hit prediction: Predicting billboard hits using spotify data. arXiv preprint arXiv:1908.08609 (2019)

\bibitem{10 Kim and Oh}
Kim, S.T. and Oh, J.H.: Music intelligence: Granular data and prediction of top ten hit songs. Decision Support Systems, \textbf{145} 113535 (2021)

\bibitem{11 Fan and Casey}
Fan, J. and Casey, M.: Study of Chinese and UK hit songs prediction. In Proceedings of the International Symposium on Computer Music Multidisciplinary Research, 640-652 (2013)

\bibitem{12 Song Popularity Predictor Homepage}
Song Popularity Predictor Homepage, \url{https://towardsdatascience.com/song-popularity-predictor-1ef69735e380}.  last accessed 17th October 2021

\bibitem{13 Kawawa-Beaudan and Garza}
Kawawa-Beaudan, J. and Garza, G.: Predicting Billboard Top 100 Songs (2015)

\bibitem{14 Borg et al.}
Borg, N. and Hokkanen, G.: What makes for a hit pop song? What makes for a pop song. Unpublished thesis, Stanford University, California, USA (2011)

\bibitem{15 Zangerle et al.}
Zangerle, E., Vötter, M., Huber, R. and Yang, Y. H.: Hit Song Prediction: Leveraging Low-and High-Level Audio Features. In ISMIR 319-326 (2019)

\bibitem{16 Essa et al.}
Essa, Y., Usman, A., Garg, T. and Singh, M. K.: Predicting the Song Popularity Using Machine Learning Algorithm (2022)

\bibitem{17 Singhi and Brown}
Singhi, A. and Brown, D. G.: Can song lyrics predict hits. In Proceedings of the 11th International Symposium on Computer Music Multidisciplinary Research 457-471 (2015) 

\bibitem{18 Chawla et al.}
Chawla, N. V., Bowyer, K. W., Hall, L. O. and Kegelmeyer, W. P.: SMOTE: synthetic minority over-sampling technique. Journal of artificial intelligence research \textbf{16}, 321-357 (2002)

\bibitem{19 Kinga}
Kinga, S.: The attributes and values of folk and popular songs. Journal of Bhutan Studies (2001)

\bibitem{20 Han et al.}
Han, J., Kamber, M. and Pei, J.: Data Mining: Concepts and Techniques (3rd ed.). Elsevier Inc (2012)

\end{thebibliography}
%

\end{document}